# Scoping Software Process Models - Initial Concepts and Experience from Defining Space Standards

Ove Armbrust[1], Masafumi Katahira[2], Yuko Miyamoto[3], Jürgen Münch[1], Haruka Nakao[4], Alexis Ocampo[1]

[1]Fraunhofer IESE, Fraunhofer-Platz 1, 67663 Kaiserslautern, Germany
[2,3]Japanese Aerospace Exploration Agency, 2-1-1 Sengen, Tsukuba, Ibaraki, 305-8505 Japan
[4]Japan Manned Space Systems Corporation, 1-1-26, Kawaguchi, Tsuchiura, Ibaraki, 300-0033
[1]{armbrust, ocampo, muench}@iese.fraunhofer.de
[2]katahira@computer.org
[3]miyamoto.yuko@jaxa.jp
[4] haruka@jamss.co.jp

**Abstract.** Defining process standards by integrating, harmonizing, and standardizing heterogeneous and often implicit processes is an important task, especially for large development organizations. However, many challenges exist, such as limiting the scope of process standards, coping with different levels of process model abstraction, and identifying relevant process variabilities to be included in the standard. On the one hand, eliminating process variability by building more abstract models with higher degrees of interpretation has many disadvantages, such as less control over the process. Integrating all kinds of variability, on the other hand, leads to high process deployment costs. This article describes requirements and concepts for determining the scope of process standards based on a characterization of the potential products to be produced in the future, the projects expected for the future, and the respective process capabilities needed. In addition, the article sketches experience from determining the scope of space process standards for satellite software development. Finally, related work with respect to process model scoping, conclusions, and an outlook on future work are presented.

## 1 Introduction

Many facets of process technology and standards are available in industry and academia, but in practice, significant problems with processes and process management remain. Rombach [1] reports a variety of reasons for this: Some approaches are too generic, some are too specific and address only a small part of daily life. Many approaches are hard to tailor to an organization's needs. In addition, some approaches impose rather strict rules upon an organization – but since not everything can be foreseen, there must be room for flexibility. Yet it remains unclear what must be regulated, and what should be left open. In general, support for process problems is plentiful, but very scattered, without a systematic concept addressing problems in a comprehensive way. One result of this unsatisfactory support is an

2      Ove Armbrust1, Masafumi Katahira2, Yuko Miyamoto3, Jürgen Münch1, Haruka Nakao4, Alexis Ocampo1

unnecessarily high number of process variants within an organization. For example, each department of a company may have its own process variant, all of them varying only in details, but existing nevertheless and needing to be maintained in parallel.

A traditional countermeasure taken to overcome this phenomenon is to define fixed process reference standards like the German V-Modell® XT [2], which fulfill the requirements of maturity models such as CMMI or ISO/IEC 15504 [3]. While this potentially reduces the number of variants, it often also leads to very generic processes that are no great help in dealing with daily problems, and that do not provide the necessary variability for coping with changing contexts. Thus, processes and their variations must be modeled in order to be understood, but at the same time, the modeling extent must be limited, in order to maintain high quality of the modeled processes and achieve high user acceptance.

Together, these circumstances have gradually turned software process management into a complex problem – and this process is nowhere near finished. As a consequence, software process management challenges comprise, but are not limited to, the following key issues:

− How can processes be characterized? Such a characterization is necessary in order to decide which process parts should become mandatory, variable, or be left out completely.
− How can stable and anticipated variable process parts be identified?
− In order to account for unanticipated changes, process models must be variable to some extent – but what is the right degree of variability?
− How can variable processes be adequately described in a process model?
− How can process models be tailored efficiently, based on the particular demand?
− On which level(s) of granularity should process engineers work?

We propose a systematic approach of Software Process Scoping to address these questions. We define Software Process Scoping as the *systematic characterization of products, projects, and processes and the subsequent selection of processes and process elements, so that product development and project execution are supported efficiently and process management effort is minimized*.

This paper is structured as follows. Section 2 names a number of requirements that a Software Process Scoping approach addressing the issues listed above should satisfy. Section 3 explains our initial solution, followed by the description of its application at JAXA in Section 4. We give an overview of what process scoping means in the aerospace domain and describe our experiences. Related work and its relationship to Software Process Scoping are analyzed in Section 5. Finally, we draw some conclusions and give an outlook in Section 6.

## 2 Requirements for Software Process Scoping

Based on the problems observed with software process management, we have phrased a number of requirements for an approach to scoping software processes.

(1) First of all, the approach should **support software product development** by providing an appropriate selection of necessary processes. This means that for a collection of existing, planned, and potential products developed in specific projects, the approach should determine the extent and provide a selection of software development processes that supports the creation of these products by providing, for each product and project, the processes needed.

(2) Second, in order to support the selection process, the approach should provide ways to **characterize software products, projects, and processes** accordingly. Since the approach is supposed to provide tailored processes for an organization, it must also provide ways to select these processes, based on process characteristics and the specific needs of projects and (future) products to be developed.

(3) Third, in order to minimize process management effort, the approach should provide ways to **distinguish stable process parts from variable ones.** Many products and projects often share large parts of the processes, with none or only minimal variations. Managing all these variants independently significantly increases process management effort. Therefore, the approach should identify stable and variable process parts, and provide a systematic method for classifying process parts accordingly, so that process management effort can be effectively decreased.

(4) Fourth, in order to cope with the unforeseen, the approach should provide ways to **incorporate unanticipated variability in a controlled manner,** such as process changes during project runtime. This requirement comes from the fact that usually, not all events can be foreseen, and thus need to be taken care of as they occur. In some cases, this requires process changes. The approach should support these changes in such a way that it sensibly constrains and guides process changes after the start of a project.

As necessary preconditions for Software Process Scoping, the following two requirements must also be fulfilled by the process modeling mechanisms used:

(1) The process modeling approach should provide ways to **store stable and variable parts within one process model,** in order to facilitate model management. Obviously, the information about whether a process part is stable or variable, and the variability's further circumstances must be stored somehow. We suggest storage within one combined model in order to facilitate further support, e.g., through tools.

(2) The process modeling approach should provide ways to **cost-efficiently instantiate such a combined model into a project-specific process model** without variability. This means that the combined model is transformed, and during this transformation, all variabilities are solved, resulting in a single process model without any remaining variability.

These requirements are by no means complete. They may need to be amended, refined or changed – however, they seem to be a good starting point to venture further. In the following section, we will present an initial solution that at least partially satisfies these requirements.

## 3 Initial Solution

One possible solution addressing the requirements mentioned is the concept of a *software process line* (see Fig. 1): *Scoping* determines the members of such a process line, *process domain engineering* constructs a process repository containing all stable

4      Ove Armbrust1, Masafumi Katahira2, Yuko Miyamoto3, Jürgen Münch1, Haruka Nakao4, Alexis Ocampo1

and variable process parts as well as a decision model governing when to use which variant. *Process line instantiation* extracts from the process repository one specific process instance without variability for each project, which can then be further adapted during *customization*. These activities are supported by a number of approaches, such as *software process commonality analysis [4], process model difference analysis* [5], [6]) and *rationale support for process evolution ([7], [8])*. In this software process line environment, scoping and process domain engineering pro-actively cope with stable and anticipated variable processes, while customization (often also just called "process tailoring") re-actively integrates unanticipated variability into a descriptive process model.

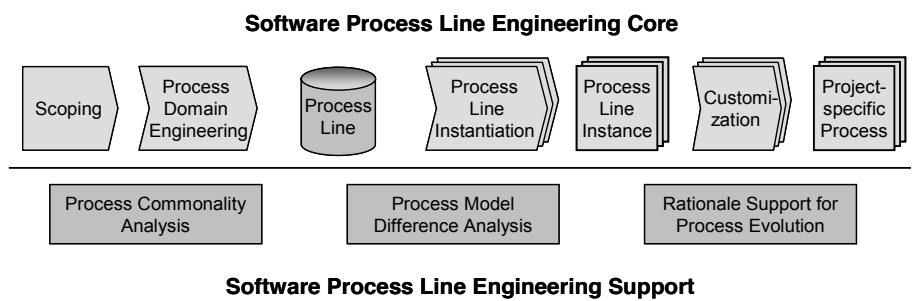

**Fig. 1.** Software process line overview

The fundamental difference between this software process line concept and well-known concepts of software process tailoring is that within a software process line, an organization's processes are actively prepared for a number of anticipated needs beforehand and then possibly tailored further to incorporate unanticipated changes, whereas classic process tailoring typically modifies a process individually for a specific project, e.g., for the creation of product P1 in cooperation with suppliers S1 and S2, to be delivered to customer C1. Within a process line, scoping would evaluate how many products of the product P1 type are anticipated to be produced in the future, how often cooperation with suppliers S1 and S2 would presumably occur, and how many projects with customer C1 are likely to happen. Taking this into account, scoping then defines mandatory and optional process parts as determined by the results of the evaluation, and process domain engineering provides the appropriate process model which reflects the scoping results.

The software *process* line concept is in fact quite similar to software *product* lines: In a product line, a software product is systematically prepared to suit future anticipated needs by determining a common core and various variants satisfying different specific needs. The software process line concept transfers this idea to software processes in such a way that it prepares an organization's processes to suit future anticipated (process) needs by determining a common process core and variable process parts that satisfy specific needs. Since product creation and processes have a very close relationship, a combination of both approaches seems only too reasonable, and was, in fact, already envisioned by Rombach in [1].

In this article, we focus on the concept of scoping. Scoping determines what to include in the process line and what not, based on characteristic features described in product, project, and process maps.

Product characteristics determine which process capabilities are needed to develop the respective product(s). These characteristics may include, for example, that a product is safety-critical, or that its requirements are only vaguely known at the beginning of development. Product characteristics are determined in a *product map* for current, future, and potential products.

Project characteristics also influence which process capabilities are needed in development projects. Such characteristics may be, for example, that a project must follow a certain development standard, or that it is performed in a distributed manner. Project characteristics are recorded in a *project map* for existing/historical, future, and potential projects. Both product and project characteristics may be prioritized, for example by their likelihood of really becoming necessary, or by the potential damage that may occur if they become necessary, but are not considered in the company's processes.

Once the product and project characterizations are complete, they form a set of demands for the company's processes. For example, product characterization may lead to the insight that for every product, its certification according to a certain standard is mandatory. Project characterization may reveal that in all upcoming projects, SPICE compliance is a must, while support for distributed development is only needed for some projects. Together, these results demand processes that are SPICE-compliant and allow for the necessary product certification by default, while explicit support for distributed development is less important.

**Fig. 2.** Product, project, and process map sketches

Available processes are characterized using the same attributes, describing the capabilities of a process in a *process map* for existing, future, and potential processes and thus providing the counterpart to the demands of products and projects. By matching the prioritized product and project characteristics to the process characteristics, the scope of the future company processes is determined, with "must have"-process features being part of the standard process and optional features representing capabilities needed only in some cases. In our simple example, SPICE-compliant processes that also support the desired certification would be included as a core for every development project, while explicit support for distributed development would be an optional feature that can be invoked on demand. Capabilities needed only very seldom or never are left out; in case they become necessary, the project-specific process tailoring will supply them.

Fig. 2 shows sketches of the three tables explained above. The topmost table displays product characteristics for existing, future, and potential products, with future products being concretely planned and potential products being a possibility, but not yet devised in any way. The middle and bottom tables contain project and process characteristics, featuring the same three-way distinction, where existing processes are processes in daily use, future processes are processes that have been prepared for application, but have not been institutionalized yet, and potential processes are processes that might become used, but have not been adapted or prepared for use within the organization yet.

## 4 Case Study

In this section, we are reporting on our experiences from an ongoing effort within the Japanese Space Exploration Agency (JAXA) to provide a process line for their space software development. Our focus hereby lies on scoping for satellite software development (see Fig. 3). In the next section, we will describe the project context and results. Following up on that, we will share our experiences.

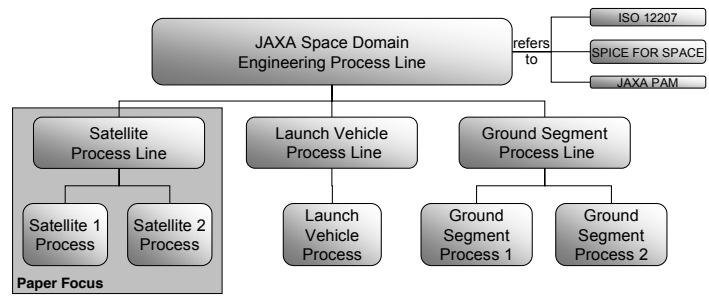

**Fig. 3.**  JAXA process line overview

### 4.1 Process Scoping in the Aerospace Domain

The ultimate goal of the ongoing project we are reporting on is to provide a software process line for JAXA's space software development. This includes satellite software, launch vehicle software, and ground segment software (see Fig. 3). So far, the first version of a satellite software process line has been finished, the scoping portion of which provided characterizations of two products (satellites) developed in two projects. In this domain, there is a very strong correlation between product and projects, since each product is unique. Nevertheless, a meaningful project and product characterization is not trivial. In our case, it became apparent very soon that while attributes for project characterization often had only two possible values (e.g., "National" and "International" for the "Collaboration type" attribute), this was not the case for product characterization. For example, complexity, criticality, and size were determined on a 3-piece scale by experts.

Tables 1 and 2 show an extract of the characterizations of the projects and products, respectively. So far, only satellite products and projects have been characterized: however, similar work for the launch vehicle and ground segment is currently going on. Due to confidentiality reasons, subsystems and suppliers are represented by numbers. In Table 2, higher numbers mean higher rating of the respective attribute.

|  |  | Satellites | | Launch Vehicle | | Ground Segment | |
|---|---|---|---|---|---|---|---|
|  |  | Sat1 | Sat2 | LV1 | LV2 | GS1 | GS2 |
| Project Characteristics | Collaboration type | National | International |  |  |  |  |
|  | Mission type | Engineering | Science |  |  |  |  |
|  | Subsystem | 1, 2, 3 | 3 |  |  |  |  |
|  | Supplier | 1, 2 | 1 |  |  |  |  |
|  | … |  |  |  |  |  |  |

**Table 1.** Excerpt from project characterization

|  |  | Satellites | | | | Launch Vehicle | | Ground Segment | |
|---|---|---|---|---|---|---|---|---|---|
|  |  | Sat1 | | | Sat2 | | | | |
|  |  | Subsystem1 | Subsystem2 | Subsystem3 | Subsystem3 | LV1 | LV2 | GS1 | GS2 |
| Product Characteristics | Complexity | 3 | 2 | 1 | 1 |  |  |  |  |
|  | Criticality | 2 | 3 | 1 | 1 |  |  |  |  |
|  | Size | 3 | 3 | 2 | 2 |  |  |  |  |
|  | Stable Requirements | yes | yes | yes | no |  |  |  |  |
|  | … |  |  |  |  |  |  |  |  |

**Table 2.** Excerpt from product characterization

There are a number of interdependencies between project and product characterization data that are not apparent at first sight, but that surfaced during scoping efforts. For example, the unstable requirements for Sat2, Subsystem3 require an iterative development approach – this led to the fact that for each potential supplier, it had to be checked whether such a process could be supported. In our case, Supplier 1 was chosen and had to adapt (for Sat2) their processes to the international collaboration type. Other interdependencies led to conflicts, e.g., the collaboration type "international" demanded that documentation had to be made available in English upon request, suggesting one set of potential suppliers, but the mission type suggested a different set – this was solved by prioritizing characteristics.

### 4.2 Experiences

Translating the project and product characterizations into requirements for the process proved not to be an easy task. Most "soft" product characteristics such as complexity, size, or criticality could not be used to directly derive new or changed processes. In fact, these factors mostly did not lead to qualitative process changes (i.e., new or changed activities or work products), but influenced project planning in such a way that the number of reviews was increased, or that the amount of independent V&V was increased. This was not modeled in detail in the software process line: instead, only high-level directives and quality requirements were given, which have to be implemented individually by the suppliers.

Project characterization, on the other hand, led to a number of variation points within the process itself. While some findings did not change the process itself (e.g., the requirement that for international projects, documentation was to be produced in English upon request), others did. For example, for international cooperation projects with ESA, a new activity was introduced for analyzing hardware/software interaction, producing the new work product FMECA (Failure Mode, Effects, and Criticality Analysis). Especially for exploratory science projects, the usual process standard was perceived as being too heavy. As a consequence, the number of quality assurance activities was reduced, and the requirements and design rationales were waived. Also, source code quality assurance measures were decreased for this type of project.

The variations were modeled using the graphical software process modeling tool SPEARMINT™. [9] Process parts that were optional in some cases were marked accordingly, with a detailed description of when to consider the respective part. Fig. 4 displays the result: The characterization information was used to derive the satellite-specific process line from the generic JAXA Space Domain Engineering Process Line. It contained a number of variable parts, the work products FMECA and Rationale for Design being shown. The rules describing these optional parts are as follows:

(Opt1.1) *if (collaboration type == international) then (produce FMECA)*
(Opt1.2) *resolve (Opt7)*
(Opt2.1) *if (mission type == engineering) then (produce Rationale for Design)*

For Opt1, two rules were defined: one that governs the creation of the FMECA, and one requiring resolution of variation point Opt7, which is concerned with the activities creating the FMECA work product (not shown). For Opt2, one rule was sufficient. Using these rules, the satellite process line could then be instantiated into two specific satellite processes. From one of these, the formerly optional parts were erased, whereas in the other one, these parts were now mandatory: The resulting Satellite 1 Process supports a national science-type project, the Satellite 2 Process an international engineering-type project.

The resulting process model contains 76 modeled activities, 54 artifacts, 18 graphical views depicting product flow, and another 18 graphical views depicting control flow. Transferring the new process model into daily practice, however, has proved to be no simple task. The modification of standards in the aerospace domain cannot be done on-the-fly because many stakeholders are involved and the consequences of software failures (possibly stemming from a faulty standard) are potentially grave. So far, the software process line we have developed has been published as an appendix to the official JAXA software standard. It has therefore not yet replaced the current standard, but JAXA engineers and their suppliers are encouraged to examine the process line and to provide comments and feedback.

Our experiences with the scoping approach taken were positive. From interviews with JAXA process engineers, we have feedback that our scoping approach helped them to focus on the relevant processes and saved a significant amount of effort in later modeling and standardization phases. The classic approach would have developed two independent processes for satellite development, so with the process line, the expected maintenance complexity has been decreased as well due to the fact that only the variable parts have to be considered separately, while for most of the process line, there is only one process to be maintained instead of two.

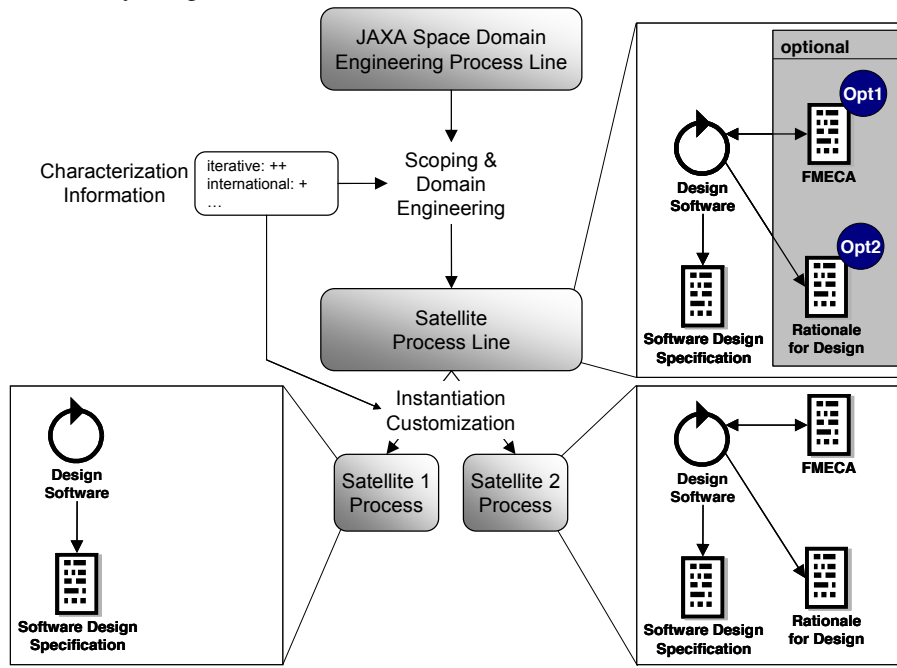

**Fig. 4.** JAXA satellite process line architecture (excerpt)

Regarding the requirements formulated before, we can state that the first three requirements concerning Software Process Scoping are already well addressed by the approach. The fourth requirement (incorporate unanticipated variability in a controlled manner) has not been addressed yet so far, which we accredit to the short lifetime of the process line: There just were no unanticipated process changes necessary yet. Considering the first process modeling mechanism requirement (storage of stable and variable parts within one process model), the JAXA project has shown that it is not feasible for larger process models to manage variable parts manually. Therefore, the tool used has been enhanced, so that it supports the definition and display of variable process elements. The second requirement (cost-efficiently instantiate a combined common/variable model into a project-specific process model), however, has not been addressed at all. JAXA did not want to provide a number of process model variants to its engineers, but instead opted for one combined model highlighting all variable parts, and describing within the model when

10      Ove Armbrust1, Masafumi Katahira2, Yuko Miyamoto3, Jürgen Münch1, Haruka Nakao4, Alexis Ocampo1

to follow which variant. This was feasible for the relatively low number of variants; however, we expect that for larger process models with more complex variants, instantiations will become necessary.

## 5 Related Work

In this section, we connect some related work to the issue of Software Process Scoping. As a basis for all scoping activities, descriptive process modeling [10] is necessary for identifying essential process entities. Becker describes an 8-step approach to descriptive process modeling. During the first step, the objectives and scope of the modeling effort are determined. This narrows the extent of the model, but the approach considers only solitary process instances on the project level, not a set of processes with variabilities. Nevertheless, descriptive process modeling can be used to determine isolated, real processes that can be used as input for a variant analysis.

Bella et al. [11] describe their approach to defining software processes for a new domain. Based on a reference process model, they used descriptive process modeling to document the as-is processes and utilized this model as a basis for deriving suitable processes for engineering wireless Internet services. Through a number of iterations, they collected qualitative and quantitative experience and adapted the processes where necessary. Their focus thus was the past; they evaluated only past events and processes. Software Process Scoping also considers the future in terms of expected products and projects.

The idea of systematically combining software product lines with matching processes was described by Rombach [1]. We consider Software Process Scoping as one potential building block of such a combined approach.

Characterization and customization approaches exist for a number of software engineering concepts, for example, for inspections [12], [13], . However, they are constrained to characterizing a limited number of methods of a class of methods (in the above case, the class of inspection methods). This comprises only a fraction of a Software Process Scoping approach, namely, that when scoping determines the need for certain characteristic features in an inspection approach, the above characterization can be used to determine which inspection approach should be used.

Denger [14] broadens the scope to quality assurance activities in general and provides a framework for customizing generic approaches to the specific needs of a company. The goal of the framework, however, is to optimize only a single factor (software quality), whereas Software Process Scoping as proposed in this article aims at optimizing multiple factors, which can be chosen freely through the product and project characterization vectors.

Avison and Wood-Harper [15] describe an approach to supply an organization with a number of methods from which a suitable one can be selected for different purposes. The authors admit that the necessary method competence for a multitude of methods is hard to achieve in reality, and therefore suggest that alternatives should be included within a single method already. Based on our experience, we support this assumption and consider this for Software Process Scoping by representing variability on different levels of abstraction.

Fitzgerald et al. [16] describe an approach taken at Motorola, which involves tailoring up-front to encompass expected deviations from the organization standard, and dynamic tailoring during project runtime, to encompass unanticipated circumstances. This corresponds to our requirements 1 and 4.

In the software product line domain, scoping has been considered in a number of publications. Clements and Northrop [17] describe three essential activities for software product line development, with scoping being a part of one of them. The authors give a detailed description of what scoping is for and what it should accomplish, but do not provide practical guidance on how to actually do it in a project. This has been done by Schmid [18]. He developed a product- and benefit-based product line scoping approach called PuLSE-Eco 2.0, which defines the scope of a software product line depending on the economical benefit of the products to be produced. The latest version of the approach is described in [19], integrating 21 customization factors that can be used to adapt the generic approach to a company's specific needs. These works were used as a basis for the Software Process Scoping approach and terminology; however, product line scoping focuses on products only and does not consider process or other context factors. Bayer et al. developed a product line based on scoping a number of business processes [20]. Their product line reflects business processes, and by determining the scope of the business processes to be implemented in software, they determined the scope of the product line. However, no more information on how scoping was done is disclosed.

Under the name of Quality Function Deployment [21], Cohen published a method for clearly specifying and ranking customer needs and then evaluating each proposed product or service capability systematically in terms of its impact on meeting those needs. This corresponds to the Software Process Scoping concepts of product/project mapping and process mapping, respectively, but is also strictly limited to products and services.

There is currently only little research going on that tries to provide a similarly systematic approach for software processes. So far, adapting processes (also known as "process tailoring") is done either generally for an organization, resulting in a single process standard, or individually for every project, resulting in a large number of process variants. Most available tailoring instructions are very generic, e.g., in international standards such as ISO/IEC 12207:1995 [3] or the German V-Modell XT [2]. However, due to their general applicability, they rarely provide more than phrases like "pick the activities and work products necessary for the purpose", and thus provide only little help in actually tailoring a process.

## 6 Conclusions and Outlook

In this paper, we presented an idea for systematically selecting and adapting software processes, depending on the project and product structure of an organization. We formulated four requirements for the approach and two requirements for supporting process modeling mechanisms, presented an initial solution addressing these requirements, and presented an application of our idea in the space software engineering domain.

Our experiences encourage us to continue on this path, and to expand the process line from satellite software development both horizontally to other branches (launch vehicle, ground segment) and vertically (JAXA-wide). The experience we collected so far supports the requirements we have set up. However, since process scoping research is yet in its infancy, a number of open questions remain. Until now, it is unclear which decision models can help to determine which process elements should be part of the process line, and which should not. A meaningful limitation of characterization attribute values (e.g., for attributes such as "complexity" or "criticality") and their objective assessment is another open issue. Furthermore, thorough investigation is needed on the subjects of how to handle different levels of abstraction in processes and characterizations (especially when talking about variability on these levels of abstraction, introduced, for example, by a vertically growing process line), how to describe interdependencies among variable parts and characterization attributes, and how to sensibly limit the number of characteristics and variation points.

Following up on what we have learned so far, our next steps will be the horizontal expansion and the inclusion of more satellite projects and products in the base of our process line, and the concurrent refinement of our approach.

## Acknowledgements

We would like to thank Ms. Sonnhild Namingha from Fraunhofer IESE for reviewing the first version of this article. We also thank the anonymous reviewers for their valuable comments on the first version of this article.